\documentclass[aps,preprint,showpacs,amsmath,amssymb,epsfig]{revtex4}
\usepackage{graphicx}
\usepackage{bm}

\begin{document}

\title{Determination of the effective strong coupling constant $\alpha_{s,g_1}(Q^2)$ from CLAS spin structure function data}

\author{
A. Deur,$^{\njlab}$
V. Burkert,$^{\njlab}$
J.P. Chen,$^{\njlab}$
W. Korsch$^{\nuk}$ 
}

\affiliation{
\baselineskip 2 pt
\centerline{{$^{\njlab}$Thomas Jefferson National Accelerator Facility, Newport
News, VA 23606}}
\centerline{{$^{\nuk}$University of Kentucky, Lexington, KY 40506}}
}

\newcommand{\njlab}{1}
\newcommand{\nuk}{2}

\begin{abstract}
 
\emph{We present a new extraction of the effective strong coupling constant 
$\alpha_{s,g_1}(Q^2)$. The result agrees with a previous determination and extends
the measurement of the low and high $Q^2$ behavior of $\alpha_{s,g_1}(Q^2)$ that was previously 
deduced from sum rules. In particular, it experimentally verifies the lack of $Q^2$-dependence 
of $\alpha_{s,g_1}(Q^2)$ in the low $Q^2$ limit. This fact is necessary for application of 
the AdS/CFT correspondence to QCD calculations. We provide a 
parameterization of $\alpha_{s,g_1}(Q^2)$ that
can equivalently be used to parameterize the $Q^2$-dependence of 
the generalized Gerasimov-Drell-Hearn and Bjorken sums.}

\end{abstract}


\maketitle

 In Quantum Chromodynamics (QCD), the gauge theory 
of the strong force, the magnitude of the coupling is given
by the running coupling constant $\alpha_{s}$. While $\alpha_{s}$ is well 
defined within perturbative QCD at large $Q^2$~\cite{BB},
these calculations lead to an infinite coupling at large distances. However, 
many  calculations, including lattice QCD or solving Dyson-Schwinger equations, indicate 
that $\alpha_{s}$ remains finite, see e.g.~\cite{review Fischer} for a 
review.
Most of these theoretical results also predict that $\alpha_{s}$ loses its scale dependence
at large distances (``freezing'' of $\alpha_{s}$) although there is still no firm 
consensus, see e.g. the new calculations~\cite{Fischer07} and~\cite{ABP} (the first result 
indicates that breaking of chiral symmetry causes $\alpha_{s}$ to be divergent at 
large distances while the second confirms previous results of the freezing of $\alpha_{s}$). 
Recently, an effective strong coupling 
constant, $\alpha_{s,g_1}$, has been extracted~\cite{alpha_g1} from experimental data on 
the Bjorken sum~\cite{JLab}. It is indicative of the freezing of $\alpha_{s,g_1}$. When 
the data are complemented with the generalized Gerasimov-Drell-Hearn (GDH) sum rule~\cite{gdh}  and 
the generalized Bjorken
sum rule~\cite{Bjorken}  predictions, the behavior of $\alpha_{s,g_1}$ can be established at any 
distance. Although the connection among the various theoretical techniques used to 
compute $\alpha_{s}$ 
at large distances is unclear, most results exhibit analog behavior and order of magnitude. 
Likewise, the connection between the experimental results reported in~\cite{alpha_g1} and 
theoretical techniques is not fully known but they display intriguing similarities. 
Remarkably, the $Q^2$ behavior of these calculations and of the data agree. 
In this paper, we present new results for $\alpha_{s,g_1}$ over an extended $Q^2$ range. We 
then propose a parameterization of $\alpha_{s,g_1}$ and finally discuss briefly the 
consequences of the behavior seen for $\alpha_{s,g_1}$.

We have extracted new data points on $\alpha_{s,g_1}$ following the procedure described 
in~\cite{alpha_g1} and based on the theoretical works of ~\cite{grunberg 1, brodsky 1}. 
The advantages and limitations of such extraction are discussed in~\cite{alpha_g1} and summarized here. Effective strong coupling constants are defined using first
order pQCD equations. As a consequence, they are process dependent. However, because they
can be related to each other using commensurate scale relations~\cite{brodsky 1}, they
can meaningfully be used for QCD predictions. The definition of effective strong coupling 
constants using first order pQCD equations makes them renormalization scheme and gauge 
independent and free of divergence at low $Q^2$. They also are analytic when 
crossing quark mass thresholds. Using the generalized Bjorken
sum rule is particularly suited to define the effective strong coupling constant 
$\alpha_{s,g_1}$. Its
simplicity makes theoretical calculations easier and it is constrained at both large and low $Q^2$ by well
established sum rules. Furthermore, there is a large amount of data available.
A present limitation of the use of effective strong coupling constants is that the connection
between various theoretical calculations of $\alpha_{s}$ at low $Q^2$ is so far
unclear due to the different approximations used. 
The relation between theoretical results and experimental extractions is also 
unclear. In ref.~\cite{alpha_g1}, we nonetheless compared all these quantities
in order to shed light on possible similarities. We will repeat the comparison 
here keeping in mind that the same precautions apply.  

\begin{figure}[ht!]
\begin{center}
\centerline{\includegraphics[scale=0.7, angle=0]{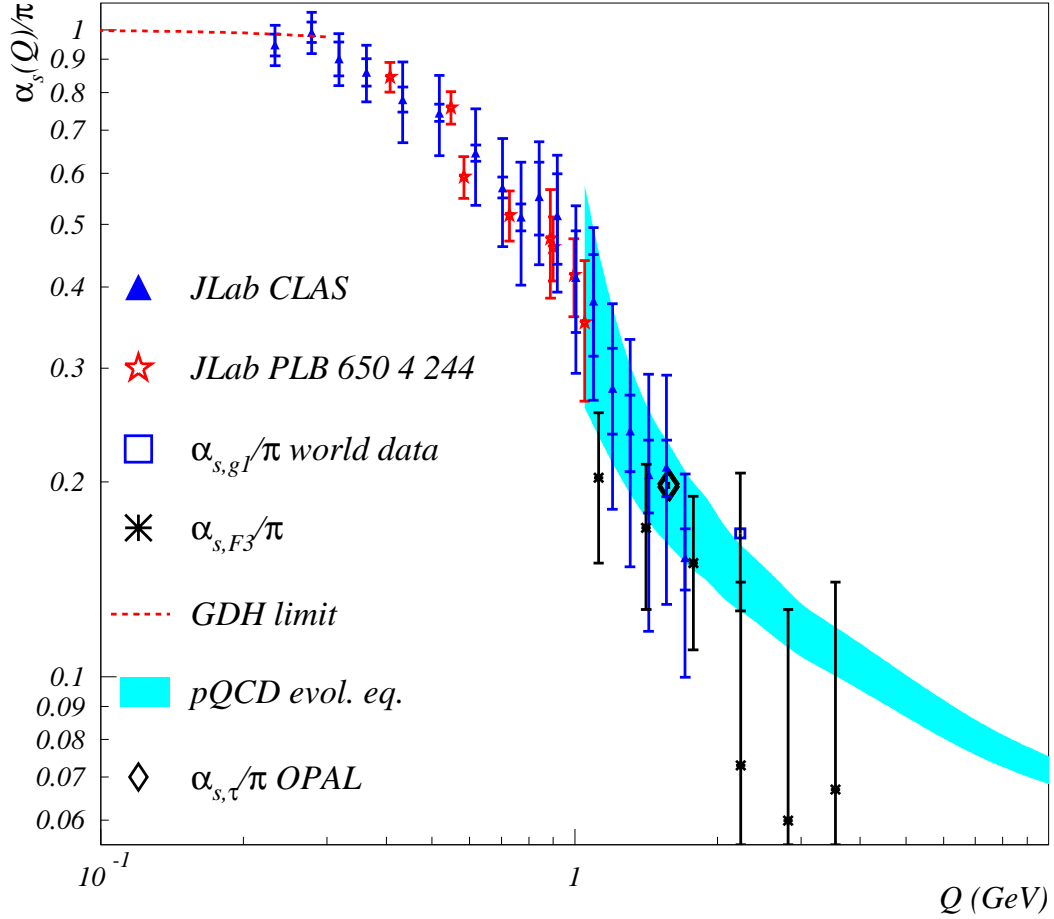}}
\end{center}
\vspace{-1cm}
\caption{(color online) $\alpha_{s,g_1}(Q)/\pi$ obtained from JLab (triangles and open stars) and world (open square) data on the Bjorken sum.  Also
shown are $\alpha_{s, \tau}(Q)/\pi$ from OPAL data, the GLS sum result from the CCFR 
collaboration (stars) and  $\alpha_{s,g_1}(Q)/\pi$ from  the Bjorken
(band) and GDH (dashed line) sum rules.}
\label{fig:alpha}
\end{figure}

The new data used to extract $\alpha_{s,g_1}$ were taken with the CLAS 
spectrometer~\cite{NIM CLAS} in Hall B at Jefferson Lab (JLab), using a 
polarized electron beam with energies ranging from 1 to 6 GeV. The data 
are reported in~\cite{EG1b} and were used to form the Bjorken sum 
$\Gamma_{1}^{p-n}(Q^2)$ in a $Q^2$-range from 0.06 to 2.92 GeV$^2$~\cite{Bjorken EG1b}. 
Here, $Q^2$ is the square of the four-momentum transfered from the electron 
to the target. Apart from the extended $Q^2$-coverage, one 
notable difference between these data and those of ref.~\cite{JLab} is that the neutron 
information originates from the longitudinally polarized deuteron target 
of CLAS while the previous data resulted from the longitudinally and transversally 
polarized $^3$He  target of JLab's Hall A~\cite{NIM A}. The effective coupling $\alpha_{s,g_1}$ 
is defined by the Bjorken sum rule expressed at first order in pQCD and at leading twist. This 
leads to the relation: 
\begin{eqnarray}
\alpha_{s,g_1}=\pi \left(1-\frac{6\Gamma_{1}^{p-n}}{g_{A}} \right)
\label{eqn:alphadef}
\end{eqnarray}
\noindent
where $g_{A}$ is the nucleon axial charge. We used Eq. \ref{eqn:alphadef} to extract 
$\alpha_{s,g_1}/\pi$. The results  
are shown in  Fig.~\ref{fig:alpha}. The inner error bars represent the statistical uncertainties 
whereas the outer ones are the quadratic sum of the statistical and systematic uncertainties. 
Also plotted in the figure are the first data on $\alpha_{s,g_1}$ from~\cite{alpha_g1} and from 
the world data of the Bjorken sum evaluated at $<Q^2>$=5 GeV$^2$, $\alpha_{s,F_3}$ from the 
Gross-Llewellyn Smith (GLS) sum rule~\cite{GLS} measured by the CCFR 
collaboration~\cite{CCFR}, and $\alpha_{s, \tau}$~\cite{brodsky 2}. See~\cite{alpha_g1} for details. The 
behavior of $\alpha_{s,g_1}$ is given near $Q^{2}=0$  by the generalized GDH 
sum rule and at large $Q^2$,  where higher twist effects are negligible, by 
the Bjorken sum rule generalized to account for pQCD radiative corrections. These predictions 
are shown by the dashed line and the band, respectively, but 
they were not used in our analysis. The width of the band is due 
to the uncertainty on $\Lambda_{QCD}$. 

The values for $\alpha_{s,g_1}$ from the new data are in good agreement with the previous JLab 
data. 
While the previous data were suggestive, the freezing of $\alpha_{s,g_1}$ at low $Q^2$ is now unambiguous and in good agreement with
the GDH sum prediction. At larger $Q^2$, the new data agree with the world data and the
results from the Bjorken sum rule at leading twist. 

We fit the data using a functional form that resembles the pQCD evolution equation 
for $\alpha_s$, with an additional term $m_{g}(Q)$ that prevents $\alpha^{fit}_{s,g_1}$
from diverging when $Q^2 \rightarrow \Lambda^2$ and another term $n(Q)$ that forces 
$\alpha^{fit}_{s,g_1}$ to  $\pi$ when $Q^2 \rightarrow 0$. Note that 
the latter constraint is a consequence of both the generalized GDH and Bjorken sum rules \cite{alpha_g1}. Our fit form is:
\begin{equation}
\alpha^{fit}_{s,g_1}=\frac{\gamma n(Q)}{log(\frac{Q^{2}+m_{g}^{2}(Q)}{\Lambda^2})}
\end{equation}
\noindent 
where $\gamma=4/\beta_{0}=12/(33-8)$, 
$n(Q)=\pi(1+[\frac{\gamma}{log(m^2/\Lambda^2)(1+Q/\Lambda)-\gamma}+(bQ)^{c}]^{-1})$ and 
$m_{g}(Q)=(m/(1+(aQ)^{d}))$. The fit is constrained by the data, the GDH and Bjorken sum 
rules at intermediate, low and large $Q^2$ respectively. 
The values of the parameters 
minimizing the $\chi ^2$ are:
$\Lambda=0.349 \pm 0.009$ GeV,
$a=3.008 \pm 0.081$ GeV$^{-1}$,
$b=1.425 \pm 0.032$ GeV$^{-1}$,
$c=0.908 \pm 0.025$,
$m=1.204  \pm 0.018$ GeV,
$d=0.840  \pm 0.051$  
for a minimal reduced $\chi^2$ of 0.84. The inclusion of the systematic uncertainties in the 
fit explains why the reduced $\chi^2$ is smaller than 1. The term $m_{g}(Q)$ has been 
interpreted within some of the Schwinger-Dyson calculations as an effective gluon 
mass~\cite{Cornwall}. Eqs. 2 and 1 can also be used to parameterize the generalized Bjorken and GDH sums.

\begin{figure}[ht!]
\begin{center}
\centerline{\includegraphics[scale=0.7, angle=0]{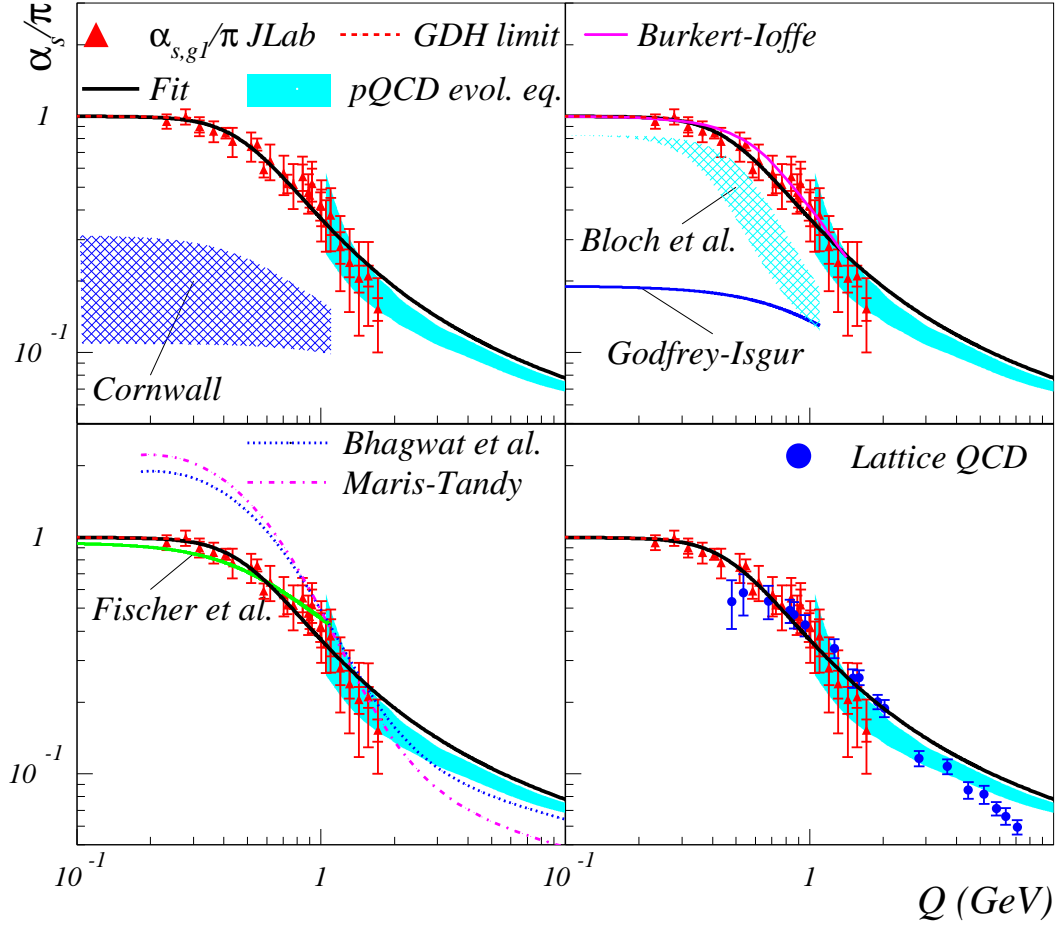}}
\end{center}
\vspace{-1cm}
\caption{(color online) The effective coupling constant $\alpha_{s,g_1}$ extracted from JLab 
data, from sum rules, and from the phenomenological model of
Burkert and Ioffe~\cite{Burkert}. The black curve is the result of the fit discussed in the text.
The calculations
on $\alpha_{s}$ are: top left panel:  Schwinger-Dyson calculations Cornwall~\cite{Cornwall}; 
top right panel: Schwinger-Dyson calculations from Bloch~\emph{et al.}~\cite{Bloch2}  
and $\alpha_{s}$ used in the quark model of Godfrey-Isgur~\cite{Godfrey-Isgur};  
bottom left: Schwinger-Dyson calculations from Maris-Tandy~\cite{Tandy2},
Fischer~\emph{et al.}~\cite{Fischer} and 
Bhagwat~\emph{et al.}~\cite{Bhagwat}; bottom right: Lattice QCD results from Furui 
and Nakajima~\cite{Furui2}.}
\label{fig:alpha2}
\end{figure}

The fit result is shown in Fig~\ref{fig:alpha2}. We also include some of the 
theoretical calculations (Lattice results and curves labeled Cornwall, Bloch \emph{et al.} and 
Fischer \emph{et al.}) and  phenomenological model predictions (Godfrey-Isgur, Bhagwat \emph{et al.} 
and Maris-Tandy) on $\alpha_s$. Finally, we show the $\alpha_{s,g_1}$ 
formed using a phenomenological model of polarized lepton scattering off polarized 
nucleons (Burkert-Ioffe). These 
calculations are discussed in~\cite{alpha_g1}. The magnitude of the 
Godfrey-Isgur and Cornwall results agrees with the estimate of the average value of 
$\alpha_s$ using magnetic and color-magnetic spin-spin interactions~\cite{Franklin}. 
We emphasize that the
relation between these results is not fully known and that they should be 
considered as indications of the behavior of $\alpha_s$ rather than strict predictions.

The data show that $\alpha_{s,g_1}$ loses its $Q^2$-dependence both at large and small $Q^2$.
The $Q^2$-scaling at large $Q^2$ is long known and is the
manifestation of the asymptotic freedom of QCD~\cite{Politzer-Gross-Wilzcek}. The absence of $Q^2$-dependence at low $Q^2$
has been conjectured and observed by many calculations but this is the first experimental 
evidence. 
This lack of scale dependence (conformal behavior) at low $Q^2$ 
shows that conformal field theories might be applicable to study the properties of hadrons.
In particular the AdS/CFT correspondence~\cite{ads/CFT} between strongly coupled gauge fields 
and weakly coupled string states can be used. This opens promising 
opportunities for calculations in the non-perturbative regime of 
QCD~\cite{ads/CFT appl 1, ads/CFT appl 2, ads/CFT appl 3}.

Finally, it is noteworthy that conformal behavior is broken in the $Q^2$-range 
between $\approx 0.7$ to a few GeV$^2$. This domain is the transition region between the 
fundamental degrees of freedom of QCD (partons) to its effective ones 
(hadrons).

To summarize, we have used new JLab data on the Bjorken sum to form the effective 
strong coupling constant $\alpha_{s,g_1}$. The $Q^2$-range is extended by factors of 3 
in both the small and large $Q^2$ sides compared to results previously reported. 
The results are in good agreement with sum rule predictions and show for the first time unambiguously
that $\alpha_{s,g_1}$ loses its $Q^2$ dependence at low $Q^2$.
We provided an analytic form for $\alpha_{s,g_1}$ (or equivalently for the 
generalized GDH and Bjorken sums) based on the pQCD result for $\alpha_{s}$
and including the sum rule constraints at small and large $Q^2$. It appears that strong 
interaction
is approximately conformal in both the large and small $Q^2$ limits. We remarked that
conformal behavior breaks down when transiting from the fundamental degrees of freedom of
QCD (quarks and gluons) to its effective ones (baryons and mesons). Establishing conformal behavior
of strong interaction is a basic step in applying the AdS/CFT correspondence to the study of 
hadronic matter.

We are would like to thanks to S. Brodsky, S. Gardner. W. Melnitchouk, C. Roberts and P. Tandy for 
helpful discussions. We acknowledge S. Furui and P. Bowman for sending us their
lattice results. 

This work was supported by the U.S. Department of Energy 
(DOE). The Jefferson Science Associates (JSA) operates the 
Thomas Jefferson National Accelerator Facility for the DOE under contract 
DE-AC05-84ER40150.

\end{document}